\documentclass[preprint]{elsarticle}
\pdfoutput=1
\usepackage{graphicx}
\usepackage{booktabs,tabu}
\usepackage[singlelinecheck=false,labelfont=bf]{caption}
\usepackage[flushleft]{threeparttable}
\usepackage{textgreek}   
\usepackage{enumitem}    
\usepackage{algorithm}   
\usepackage{algorithmic}

\makeatletter          
\def\ps@pprintTitle{%
 \let\@oddhead\@oddhead
 \let\@evenhead\@empty
 \def\@oddfoot{\leftline{}}
 \let\@evenfoot\@oddfoot}
\makeatother

\journal{}
\begin{document}
\begin{frontmatter}
\title{A heuristic approach for lactate threshold estimation for training decision-making: An accessible and easy to use solution for recreational runners}

\author[mainaddress]{Urtats Etxegarai\corref{cor1}}
 \ead{urtats.etxegarai@ehu.eus}
 \cortext[cor1]{Corresponding author.Tel.:+34 65 371 9204}
\author[mainaddress]{Eva Portillo}
\author[addressTwo]{Jon Irazusta}
\author[addressThree]{Lucien Koefoed}
\author[addressThree]{Nikola Kasabov}

\address[mainaddress]{Department of Automatic Control and System Engineering, Faculty of Engineering, University of the Basque Country UPV/EHU, Bilbao, Spain}
\address[addressTwo]{Department of Physiology, Faculty of Medicine and Nursing, University of the Basque Country UPV/EHU, Leioa, Spain}
\address[addressThree]{Knowledge Engineering and Discovery Research Institute, Auckland University of Technology, Auckland, New Zealand}

\begin{abstract}

In this work, a heuristic as operational tool to estimate the lactate threshold and to facilitate its integration into the training process of recreational runners is proposed. To do so, we formalize the principles for the lactate threshold estimation from empirical data and an iterative methodology that enables experience based learning. This strategy arises as a robust and adaptive approach to solve data analysis problems. We compare the results of the heuristic with the most commonly used protocol by making a first quantitative error analysis to show its reliability. Additionally, we provide a computational algorithm so that this quantitative analysis can be easily performed in other lactate threshold protocols. With this work, we have shown that a heuristic (\%60 of \textit{endurance running speed reserve}), serves for the same purpose of the most commonly used protocol in recreational runners, but improving its operational limitations of accessibility and consistent use.

\end{abstract}

\begin{keyword}
\texttt{OR in sports}\sep Lactate \sep Anaerobic threshold \sep Decision support systems \sep Heuristics    
\end{keyword}

\end{frontmatter}

\section{Introduction}
\label{sec:intro}
In sport performance, where the context is complex and ever changing, the main need of coaches and athletes is to discover ways of integrating as much meaningful contextual information as possible into the decision-making process.

Given that there is some meaningful information available, it is common that the constraints hindering its integration in the decision-making process are operational. In sport performance, these operational constraints are usually related to: accessibility (location, requirement of specialized equipment, excessive associated costs etc.), interference with training (sub-maximum tests, time to integration etc.) and facilitating the adherence and its consistent use (ease of use, comfort etc.) among others.

The lactate threshold (LT) is the exercise intensity at which the concentration of blood lactate begins to exponentially increase compared to the values at resting. This physiological indicator is considered essential for endurance sports performance from (at least) three different perspectives: monitoring training by assessing the physical condition from an endurance performance perspective \cite{Pallares2016}; prescribing exercise intensity \cite{Hofmann2017}; and performance estimation \cite{Pallares2016}.

However, there is currently no reliable method for assessing the LT without specialized equipment and/or personnel, meaning that it is restricted to few people with access to these resources. Moreover, the 'gold standard' for assessing the LT is via taking blood samples \cite{Santos-Concejero2014a}, which is a cumbersome and uncomfortable process inconvenient for its consistent use.

The approach of using the LT for training decision-making is very extended among endurance coaches and athletes, and actually, there is a big recreational runner population who demand an accessible training decision-making method \cite{Runningusa}. Therefore, the main objective of this work is to propose an accessible method to estimate the LT and to facilitate its integration into the training process of recreational runners.

So far, significant efforts have been made to overcome some of these operational constraints, and solutions such as non-invasive hardware (HW) \cite{Borges2015}, analytical models \cite{Proshin2013} and empirical models \cite{Erdogan2009} have been proposed. However, non of the proposed methods were able to achieve both a reasonable reliability while addressing accessibility and facilitating its consistent use. To the best of our knowledge, the model we previously proposed as a LT estimator \cite{Etxegarai2018} is the only alternative that has covered all these operational requirements. However, creating an estimator from empirical data is an iterative process of problem discovery where the complexity of the problem and its relation with the available data is further discovered with each iteration. 

In \cite{Etxegarai2018} we run a first iteration and proposed a first model as estimator, hereafter referred as the \textit{initial estimator}. The empirical LT was obtained using the Dmax LT protocol, which is the most recommended method for the LT estimation from blood samples \cite{Santos-Concejero2014a}. In this first iteration, we found that estimating the LT from empirical data is viable and elucidated the main characteristics and the degree of complexity of the problem. However, a deeper performance evaluation of both the methodology and the \textit{initial estimator} is still needed for a real applicability, especially in terms of its generalization performance. Therefore, in this work we formalize the first principles for the LT estimation from empirical data and an iterative methodology that enables experience based learning to improve the \textit{initial estimator} in terms of both the required generalization capabilities and accuracy. As part of this methodology, new ways to evaluate its applicability from multiple perspectives are proposed (generalization, strengths, limitations, comparison with the state-of-the-art, room for improvement and future steps).  

In particular, a heuristic is proposed as a simple and reliable solution for the LT estimation. This heuristic accomplishes the operational constraints related with accessibility and consistent use.

The rest of the article is organized as follows: Section \ref{sec:Criteria} sets the \textit{principles} on which the LT estimation methodology stands; Section \ref{sec:Methodology} presents the iterative methodology and creates a reliable and operational LT estimator; Section \ref{sec:adjust} discusses the implications of the proposal and leads the way for future steps and Section \ref{sec:conclusions} offers concluding remarks.

\section{Principles for the LT estimation from empirical data}
\label{sec:Criteria}

From the application perspective the validity of a system is related to its capacity to provide a successful solution for the problem in hand. 
 
For a solution valid for training decision-making, the importance resides in providing appropriate estimations for individual athletes. Additionally, the estimator must provide valid individual estimations for the majority of the represented population. In the following sections we formulate these concepts: 

\subsection{Performance for individual athletes: \textit{Individual acceptable error}}

\begin{figure}
\includegraphics[width=\columnwidth]{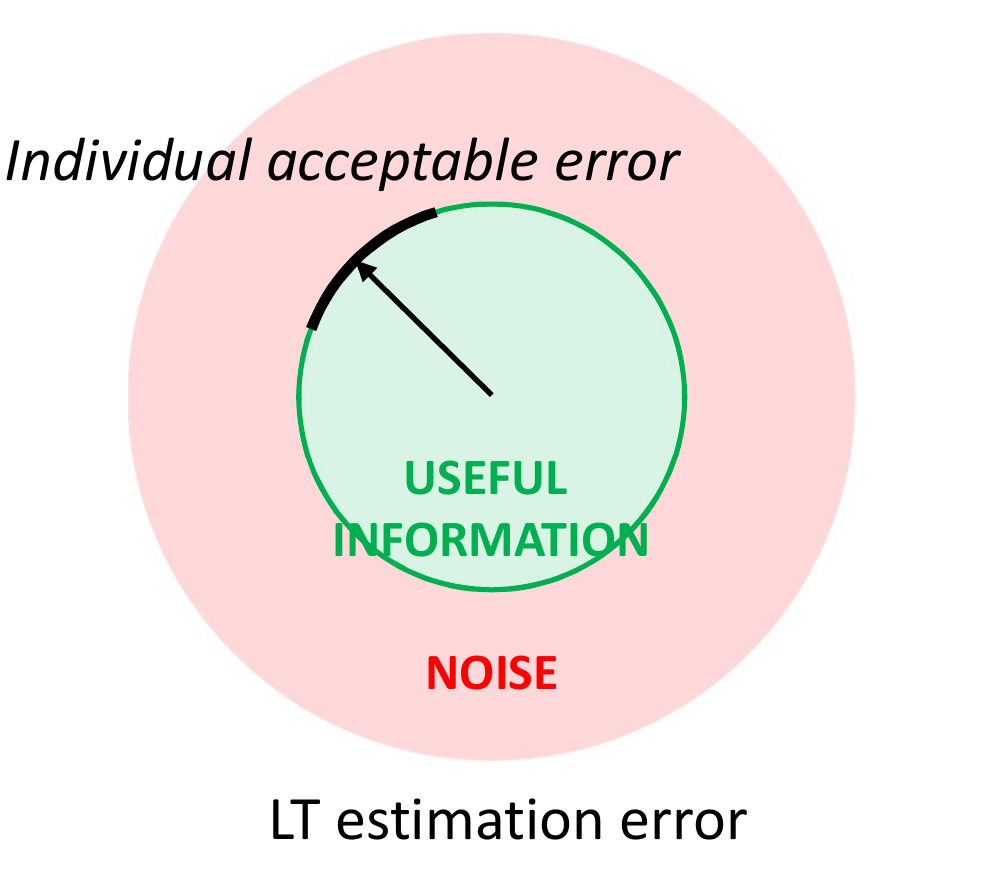}

\caption{\textit{Performance for individual athletes: \textit{Individual acceptable error}}}
\label{fig:individual}
\end{figure}

Individualization of training is key in sport performance and, to provide a tool useful in practice, the performance evaluation of the estimator must be done with this perspective in mind. In this regard, we define the \textit{individual acceptable error} as the maximum error in the LT estimation of a particular athlete so that the estimation remains useful for training decision-making. Figure \ref{fig:individual} represents this concept. In \cite{Etxegarai2018}, we proposed an \textit{individual acceptable error} for recreational athletes which we represent in table \ref{tab:SatisfyingError}.

\begin{table}
\centering
\begin{threeparttable}
\caption{Lactate threshold \textit{individual acceptable error}}
 \begin{tabular}{ccc} 
  \toprule
   Pace at the LT	& \multicolumn{2}{c}{Maximum error in the LT}\\
   (min/km) & $\pm$(s/km) & $\pm$(\%)\\
   \midrule
   $\ge$3\textsuperscript{$\ast$} & 3 & 1.7\\
   $\ge$3.5 & 5 & 2.4\\
   $\ge$4 & 10 & 4.2\\
   $\ge$4.5 & 15 & 5.5\\
   $\ge$5 & 20 & 6.6\\ 
   \bottomrule
  \end{tabular}
  \begin{tablenotes}
   \small
   \item *Out of scope: Fitness level above objective population
	 \item Abbreviations: LT, lactate threshold
	 \item Source: \cite{Etxegarai2018}
  \end{tablenotes}
   \label{tab:SatisfyingError}
\end{threeparttable}
\end{table}

\subsection{Performance for recreational runners population: \textit{System acceptable error rate} and \textit{parsimony} criteria}

The \textit{system accuracy} is the percentage of individuals that are estimated within the \textit{individual acceptable error} defined in table \ref{tab:SatisfyingError} for the represented population (i.e. recreational runners).

However, the evaluation of the \textit{system accuracy} for an estimator created from empirical data is not straightforward. The \textit{system accuracy} has to be analyzed with respect to the \textit{expected error} of the estimator, which is related with how it behaves in the underlying joint probability distribution. Therefore, the \textit{expected error} of the estimator must be below a \textit{system acceptable error rate} (1) to be useful for the majority of the recreational runner population. 

\begin{center}
expected error $\leq$ system acceptable error rate (1)
\end{center}

From the generalization perspective, there is a difference between the \textit{expected error} of an estimator and \textit{empirical error} (the error approximated using the available data). This difference is known as \textit{generalization error} (2) \cite{Abu-Mostafa2012}: 

\begin{center}
generalization error = expected error - empirical error (2)
\end{center}

Therefore, for a solution valid for the majority of the recreational runners population, the combination of \textit{empirical} and \textit{generalization error} must be below a \textit{system acceptable error rate} (3).
 
\begin{center}
empirical error + generalization error $\leq$ system acceptable error rate (3)
\end{center}

\begin{figure}
\includegraphics[width=\columnwidth]{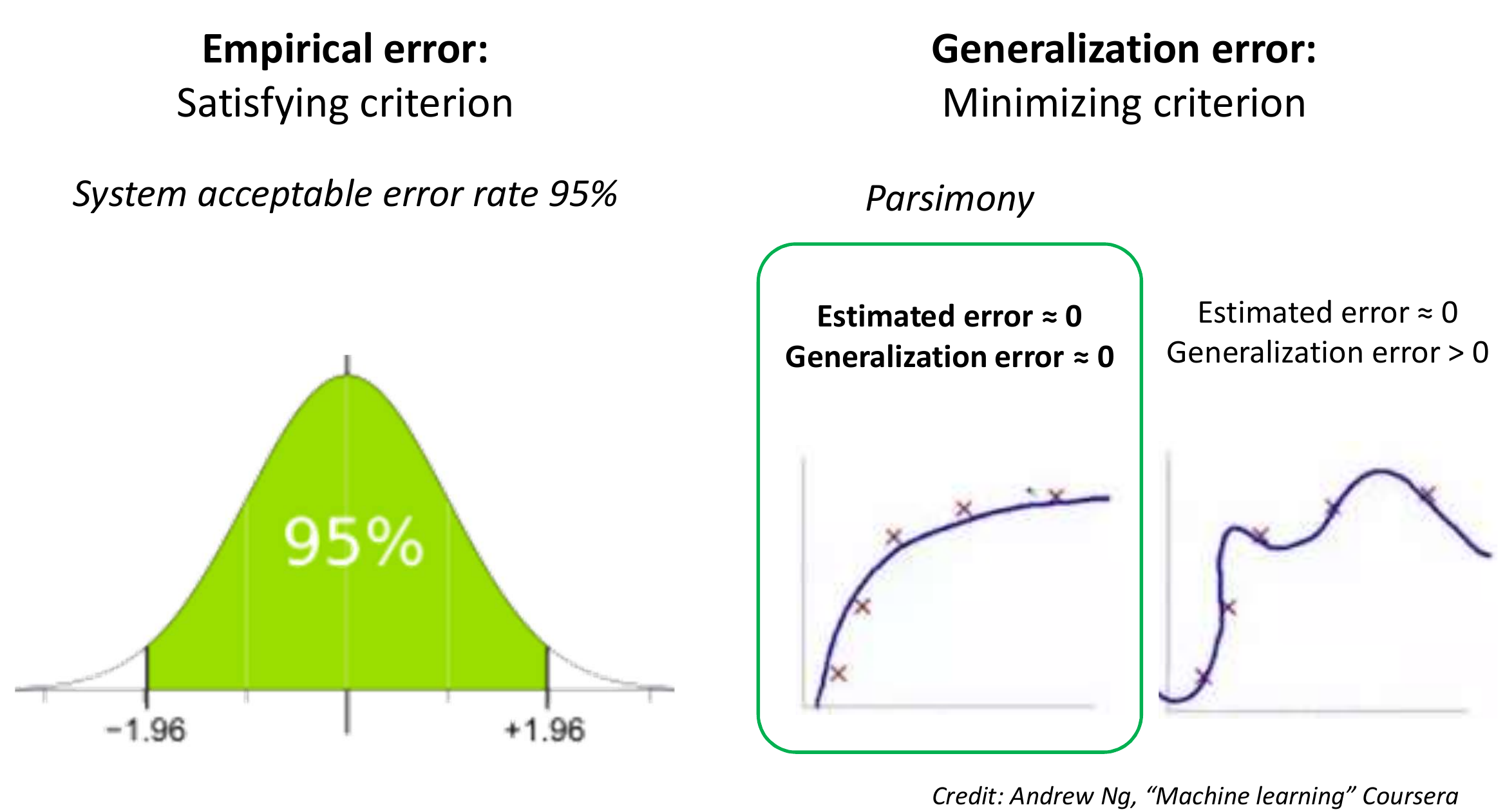}

\caption{Performance for recreational runners population: \textit{System acceptable error rate} and \textit{parsimony} criteria}
\label{fig:criteria}
\end{figure}

Since the \textit{empirical error} is observable and the \textit{generalization error} is not, we propose a quantitative satisfying criterion for the former and a qualitative minimizing criterion for the latter (figure \ref{fig:criteria}). It is important to note that, despite not being quantifiable, \textit{generalization error} can be indirectly estimated using different techniques. These techniques are an important part of the proposed methodology.

The recreational runners population, as any other heterogeneous population, has individuals that are far from what it can be considered normal. The efforts and resources required to improve the estimation power for less common athletes grow non-linearly with respect to their non-normality. Therefore, here we set the \textit{system acceptable error rate} as a satisfying criteria that sets upper bound for the \textit{empirical error}. 

The calculation of the \textit{system acceptable error rate} is done in terms of the room for improvement with respect to the ideal estimator that could be achieved with the Dmax LT protocol. For the acceptance of the estimator as a valid tool, we set the \textit{system acceptable error rate} at two standard deviations or 95\%. 

Regarding the minimizing criterion, the principle of \textit{parsimony} says that, in equal conditions, the most parsimonious approach reduces variance, making the solution more robust to unseen data comparing to other more complex ones \cite{James2000}. Moreover, \textit{parsimony} is easily assessed as it is related with the degrees of freedom that the estimator has. Thus, \textit{parsimony} is the criterion used to minimize the \textit{generalization error} of the solution. 

\section{An iterative methodology for an operational LT estimation}
\label{sec:Methodology}

\begin{figure}
\includegraphics[width=\columnwidth]{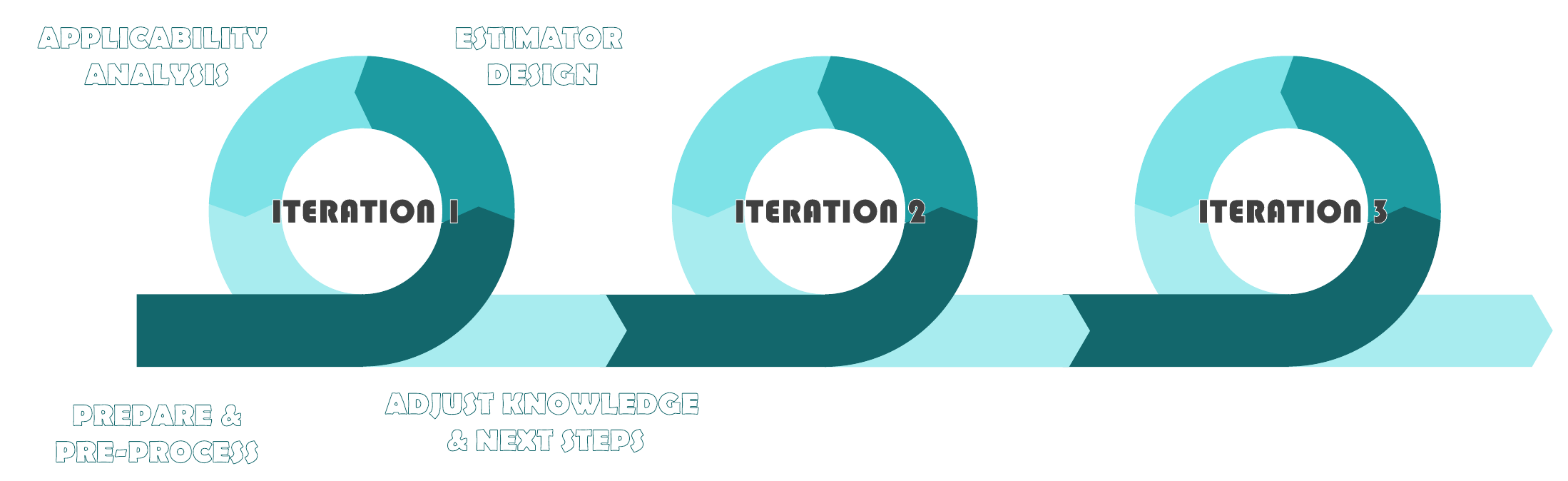}

\caption{An iterative methodology for an operational LT estimation}
\label{fig:strategy}
\end{figure}

With the principles defined in section \ref{sec:Criteria} as reference, an iterative methodology is proposed  towards an accessible, easy to use and reliable LT estimation. Figure \ref{fig:strategy} represents this approach.

This methodology is divided in four phases: preparation and pre-procesing, where, collecting new experimental data, we evaluate the limits of the initial estimator from the bias-variance perspective; estimator design, where we design an estimator according to the conclusions derived in the previous phase and the knowledge of previous iterations; applicability analysis, taking into account several perspectives such us generalization and room for improvement; and adjusting knowledge and deciding next steps, where we discuss the implications of our proposal from the methodological, sport science and training decision-making perspective as well as providing the instructions for its use. First three phases are further detailed in this section, while the fourth is presented in the next section as part of the discussion.

\begin{figure}
\includegraphics[width=\columnwidth]{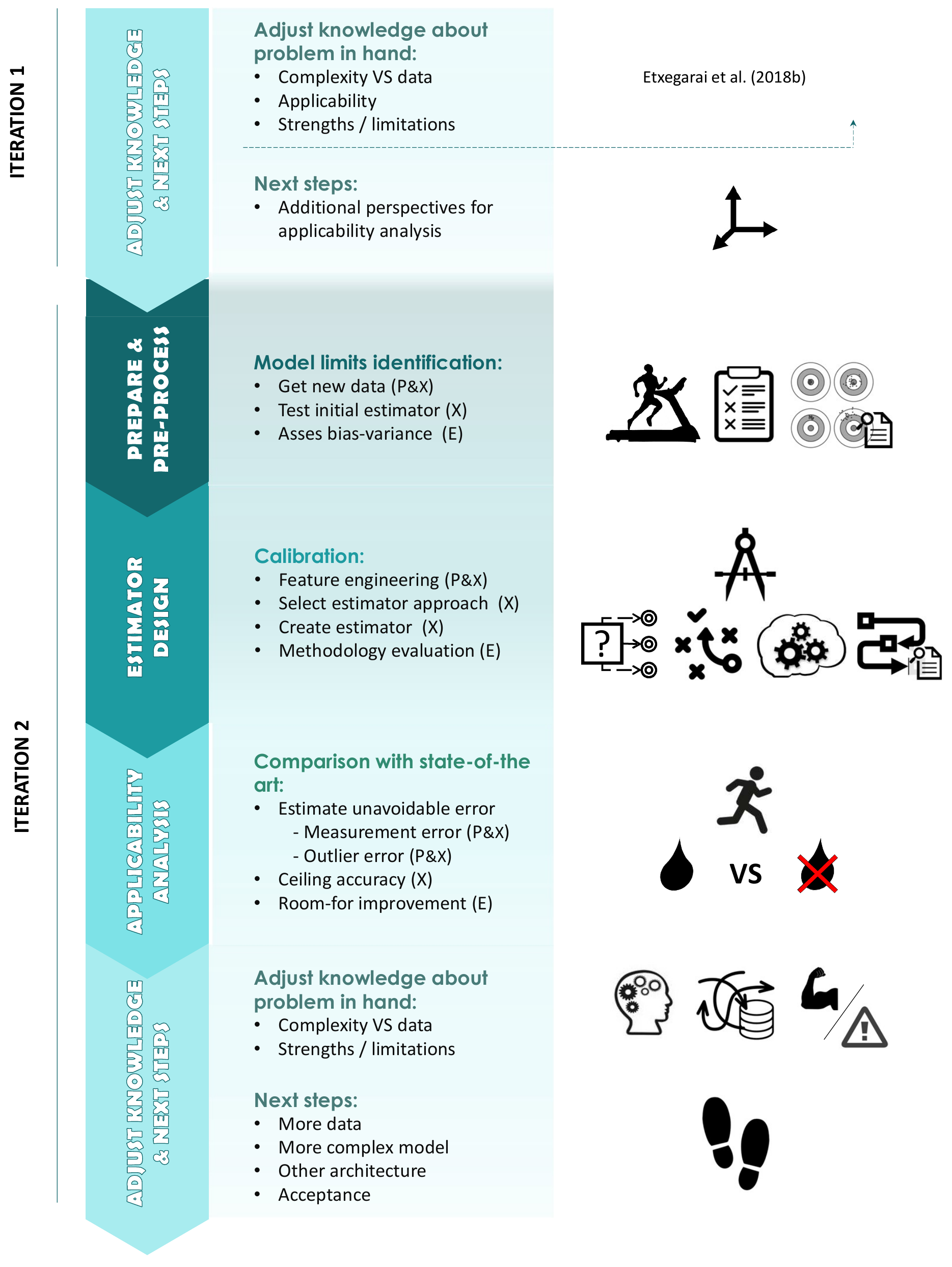}
\caption*{Abbreviations: P, planning; X, execution; E, evaluation}
\caption{Detailed methodology for 2$^{nd}$ iteration}
\label{fig:estimatorMethodology}
\end{figure}

At the same time, each phase follows a planing (P), execution (X) and evaluation (E) steps from which some conclusions are derived to serve as guide to the next phase. The detail of these four steps is represented in figure \ref{fig:estimatorMethodology} and further explained in the following paragraphs.

\subsection{Prepare \& pre-process: Model limits identification}

As we mentioned in section \ref{sec:Criteria}, the generalization error, which is closely related to the variance of the estimator, is not directly observable. There are methods by which the generalization error can be estimated, but each has drawbacks. Plotting learning curves is one possible procedure, however it has high computational cost for complex algorithms such as the Recurrent Neural Networks that we used in \cite{Etxegarai2018}. Testing the model against unseen data is another possible approach, but involves taking more experimental samples which is not always economically viable. However, it is usually better as the solution is evaluated with new and overally bigger sample, and it is also more viable in our particular case.  

Therefore, to test the limits of the initial estimator and consequently the learning, performance evaluation and estimator selection methodologies used in \cite{Etxegarai2018}, new experimental tests were made with 50 recreational athletes. The purpose of these experimental data is to serve as a test set to evaluate the initial estimator, assess its bias-variance performance and adjust the estimator accordingly in this 2$^{nd}$ iteration. The 50 athletes were randomly selected from a list of 803 athlete's that volunteered for the study. Personal interviews were also held to confirm that the selection criteria were being satisfied and that the athletes where representative of the recreational runners population. The data acquisition and preprocessing protocols follow the same steps as in \cite{Etxegarai2018} where the runners completed a maximal incremental running test at 1\% slope on a treadmill, which started at 9 km/h without previous warm up. The speed was increased by 1.5 km/h every 4 min until 13.5 km/h; then, to calculate the LT more accurately, speed was increased by 1 km/h until the participant reached exhaustion. One minute of recovery was given between stages to obtain capillary blood samples from the earlobe and measure the lactate concentration with the Lactate Pro analyzer (Arkray, KDK Corporation, Kyoto, Japan.), which is been validated as an effective analyzer \cite{Tanner2010}.

\begin{figure}
\includegraphics{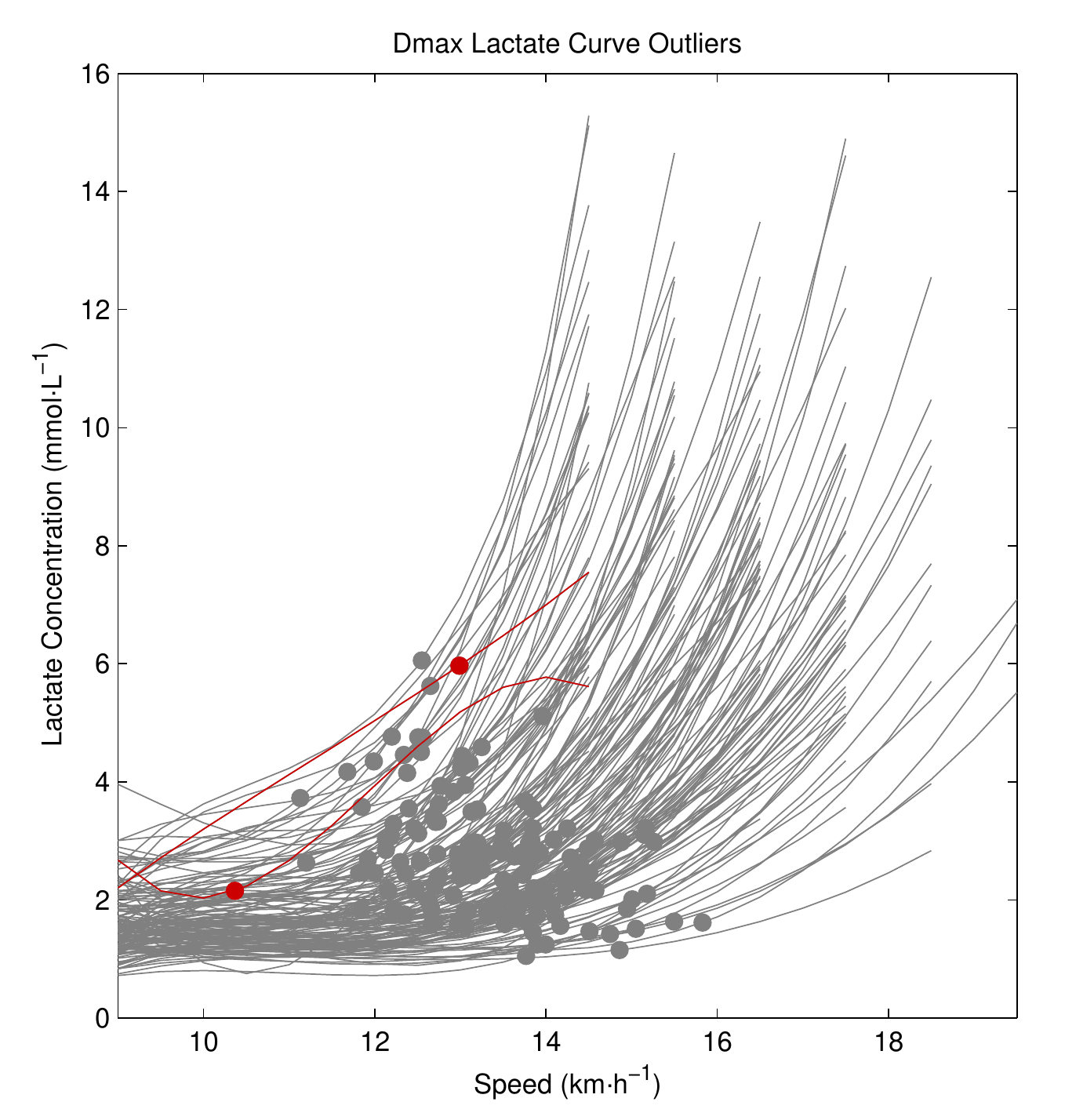}
\caption{Outlier athlete lactate curves}
\label{fig:outlier}
\end{figure}

It is well known that there are athletes for which, due to individual characteristics, the Dmax LT protocol is not suitable. In this work, we refer to these athletes as outliers. This means that, for these athletes, the Dmax LT does not represent the real LT and thus must no be used to for the estimator design nor the evaluation. Therefore, before evaluating the performance of the initial estimator, the lactate curves and the LTs of every athlete in the database are analyzed in terms of domain expertize so that the outliers are detected and removed. In figure \ref{fig:outlier}, we observe that two athletes have been detected as clear outliers.

Once that the outliers are removed, the performance of the \textit{initial estimator} is evaluated against the new experimental data and gathered in table \ref{tab:modelLimit}. As we can observe, the bias in the first iteration is small (11\% and 9\%). However, this error rate is not maintained when tested with the new database of the 2$^{nd}$ iteration (36\%) showing that the variance and consequently the \textit{generalization error} of the estimator is non-negligible. 

\begin{table}
\centering
\begin{threeparttable}
\caption{Initial estimator limits identification}
 \begin{tabular}{cccc} 
  \toprule
   Estimator & \multicolumn{2}{c}{1$^{st}$ Iter DB}	& 2$^{nd}$ Iter DB\\
    & train set perf & test set perf & test set perf\\
   \midrule
	 Initial & 89\% & 91\% & 64\%\\
   Calibrated & - & - & -\\
   \bottomrule
  \end{tabular}
  \begin{tablenotes}
   \small
	 \item Abbreviations: Iter, Iteration; DB, Data base; perf, performance
  \end{tablenotes}
   \label{tab:modelLimit}
\end{threeparttable}
\end{table}

Given these results and the general criteria defined in section \ref{sec:Criteria}, in this work we delve deeper in minimizing the \textit{generalization error} introduced in the estimator design process by following the \textit{parsimony} criterion, so that we avoid an overly optimistic \textit{empirical error} \cite{Cawley2010}.

\subsection{Estimator design: Calibration}

The calibration of the estimator design starts from the feature engineering. Feature engineering deals with the proper representation of the feature vector. As demonstrated by the big efforts placed on the design of pre-processing pipelines and data transformations \cite{Heaton2016}, a proper representation of the problem space can greatly simplify the path to the desired solution. From the \textit{parsimony} perspective, the number of features used to create an estimator is inversely related with its \textit{parsimony}. Therefore, our approach minimizes the number of features used. 

The Peak Treadmill Speed (PTS) is the maximum speed achieved in the maximum incremental running test done to gather the experimental data (i.e. the Dmax LT). In this work we define the \textit{endurance running speed reserve} as the diference between the PTS and the initial running speed of the maximum incremental running test. 

\begin{center}
endurance running speed reserve [km/h] = PTS [km/h] - initial running speed [km/h] (4)
\end{center}

In this work, a feature transformation is used to represent the LT in a way that it is decoupled from the PTS, so that it can be represented in relative terms whit respect to the individual \textit{endurance running speed reserve} (5):

\begin{center}
LT [\%] = (LT [km/h] - initial running speed [km/h]) / endurance running speed reserve [km/h] $\times$ 100 (5)
\end{center}

\begin{figure}
\centering
\includegraphics[width=\columnwidth]{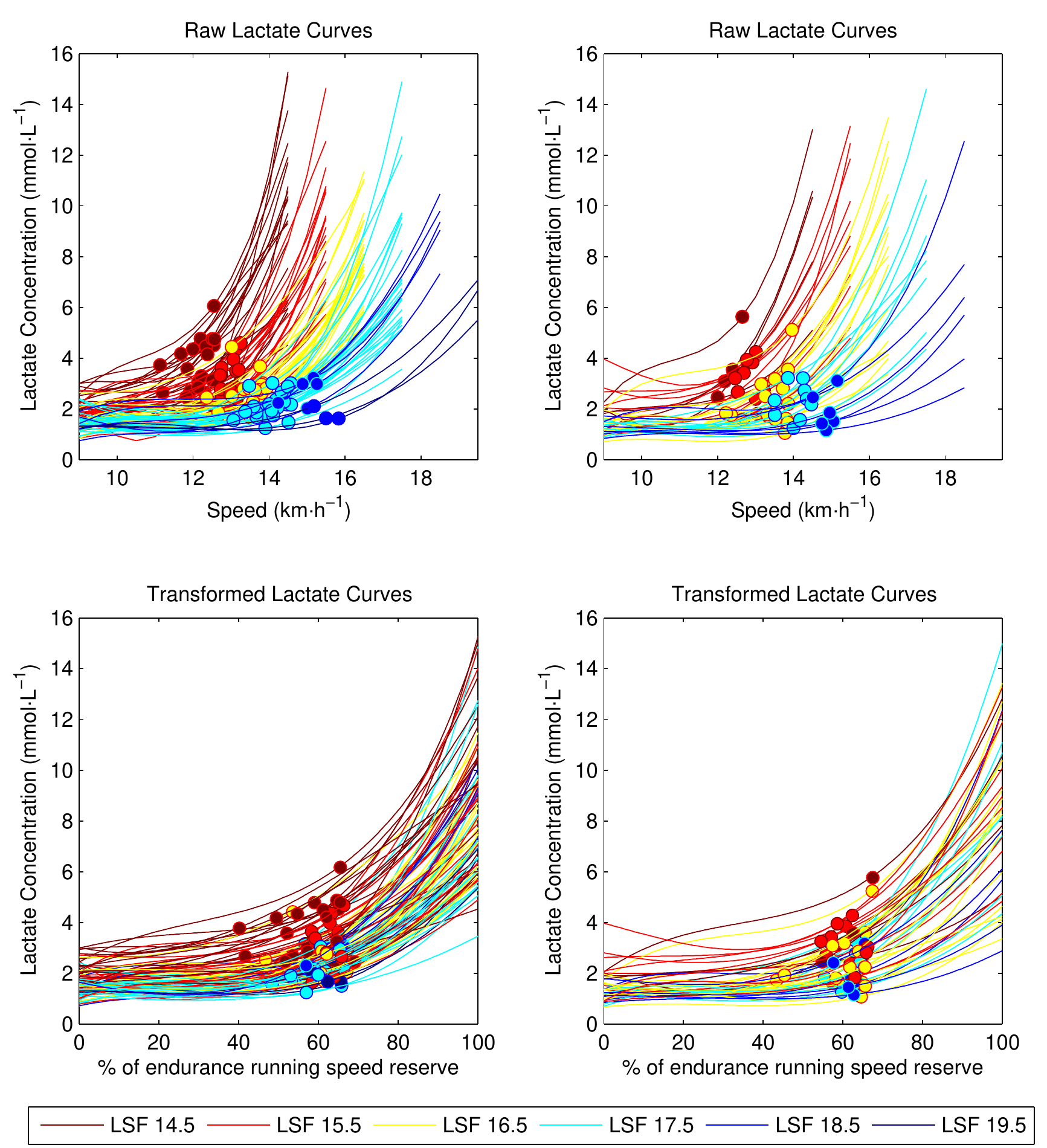}
\caption*{Abbreviations: LSF, last step finished}
\caption{Experimental lactate curves: 1$^{st}$ iteration data vs 2$^{nd}$ iteration data}
\label{fig:LactateCurves}
\end{figure}

As we highlighted in \cite{Etxegarai2018,Etxegarai2018a}, there is a consistent association between the PTS and the LT. As shown in figure \ref{fig:LactateCurves}, once we dis-attach the LT from the effect of the PTS, the variability of the LT is highly reduced and the problem is simplified.

Then, using the \% of \textit{endurance running speed reserve} corresponding to the LT as estimand, we propose to use the mean as an extremely \textit{parsimonious} estimator. It is important to note that, the mean, by definition, produces a LT estimation with no variance and therefore it is the most \textit{parsimonious} estimator that we could use. To address the confidence intervals (CI) of the mean, we have used the bias corrected and accelerated percentile bootstrap method which is a non-parametric and more robust way to estimate the CI. It shows that the CIs are close to the mean value. It is all represented in table \ref{tab:stats}. 

\begin{table}
\centering
 \begin{threeparttable}
 \caption{1$^{st}$ Iteration transformed LT's statistics}
  \begin{tabular}{@{}c@{}} 
  \toprule
  mean (boots CI)\\
  \midrule
  59.6\% (58.3 - 60.7)\\
	\bottomrule
  \end{tabular}
	\begin{tablenotes}
  \small
  \item Abbreviations: LT, lactate threshold; boots CI, bootstrap confidence interval
 \end{tablenotes}	
  \label{tab:stats}
 \end{threeparttable}
\end{table}

To evaluate the validity of the \textit{calibrated estimator} in this 2nd iteration, its performance is tested with the new experimental data. As expected, table \ref{tab:bothIter} shows very similar bias between iteration 1 and 2 datasets, due to the lack of variance of the \textit{calibrated estimator}. Moreover, the performance is still high even that we used a very \textit{parsimonious} solution.

This means that the estimator here proposed completely fulfills the \textit{parsimony} criteria established at in section \ref{sec:Criteria} and achieves \textit{system accuracy} of 87\%, almost fulfilling the \textit{system acceptable error rate} for acceptance. However, as stated in section \ref{sec:Criteria}, there is a \textit{ceiling accuracy} towards which the performance of the \textit{calibrated estimator} must be compared and that will be analyzed in the following phase.

\begin{table}
\centering
\begin{threeparttable}
\caption{Lactate threshold system acceptable error}
 \begin{tabular}{cccc} 
  \toprule
   Estimator & \multicolumn{2}{c}{1$^{st}$ Iter DB}	& 2$^{nd}$ Iter DB\\
    & train set perf & test set perf & test set perf\\
   \midrule
	 Initial & 89\% & 91\% & 64\%\\
   Calibrated & 89\% & 87\% & 87\%\\
   \bottomrule
  \end{tabular}
  \begin{tablenotes}
   \small
	 \item Abbreviations: Iter, Iteration; DB, Data base; perf, performance
  \end{tablenotes}
   \label{tab:bothIter}
\end{threeparttable}
\end{table}

\subsection{Applicability analysis: Comparison with state-of-the art}
In the previous phase, we evaluated the \textit{calibrated estimator} against the \textit{parsimony} and \textit{system acceptable error rate}. In this phase, we delve deeper into applicability of our solution, evaluating the room for improvement of it. To do so, we make a comparison between the \textit{calibrated estimator} and the Dmax LT protocol \cite{Santos-Concejero2014a}.

The Dmax LT estimation protocol, as any other methodology that tries to estimate or measure a variable, suffers from approximation errors. These errors are intrinsic to the protocol itself and set  that an empirical estimator can achieve. In this work we refer to this error as the unavoidable error. This unavoidable error set the \textit{ceiling accuracy} that and ideal empirical estimator could achieve and are used to better evaluate the the accuracy of our calibrate estimator an set its room for improvement.  

\begin{figure}
\includegraphics[width=\columnwidth]{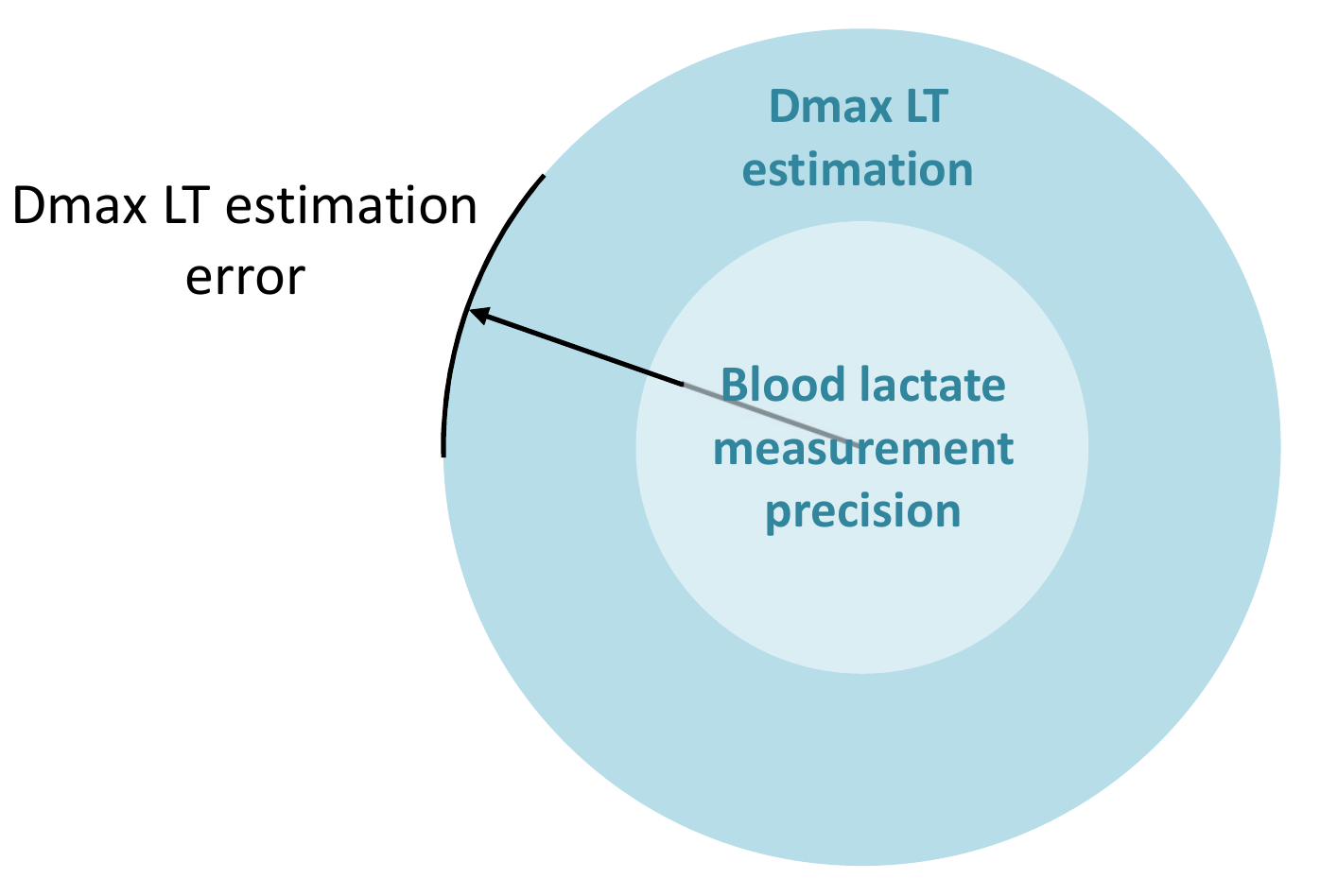}
\caption*{Abbreviations: LT, lactate threshold}
\caption{LT estimation error caused by measurement precision error}
\label{fig:irreducible}
\end{figure}

As represented in figure \ref{fig:irreducible}, the unavoidable error of the Dmax LT estimation can be caused by a combination of the blood lactate measurement error and the error that the Dmax protocol introduces when translating it into the estimated LT. The former is well characterized and quantified in \cite{Tanner2010} and is related with the measurement device error, blood sampling errors related to sweat, timing etc. The latter is also well known in the literature \cite{Santos-Concejero2014a} and it is related with the initial and final point selection, regression type, number of blood measurements etc. However, it hasn't been quantitatively addressed so far.

This error causes a random variability in the Dmax LT and, as our estimator is build on top of these empirical data, is part of the unavoidable error that our estimator has. 

To estimate this unavoidable error, we propose a computational algorithm that takes multiple random samples from the blood lactate precision distribution \cite{Tanner2010} and estimate the different plausible values of blood lactate concentration according to this error. More precisely, this computational algorithm takes random samples with replacement from the sampled population to better represent the underlying population distribution. 

\begin{figure}
\includegraphics{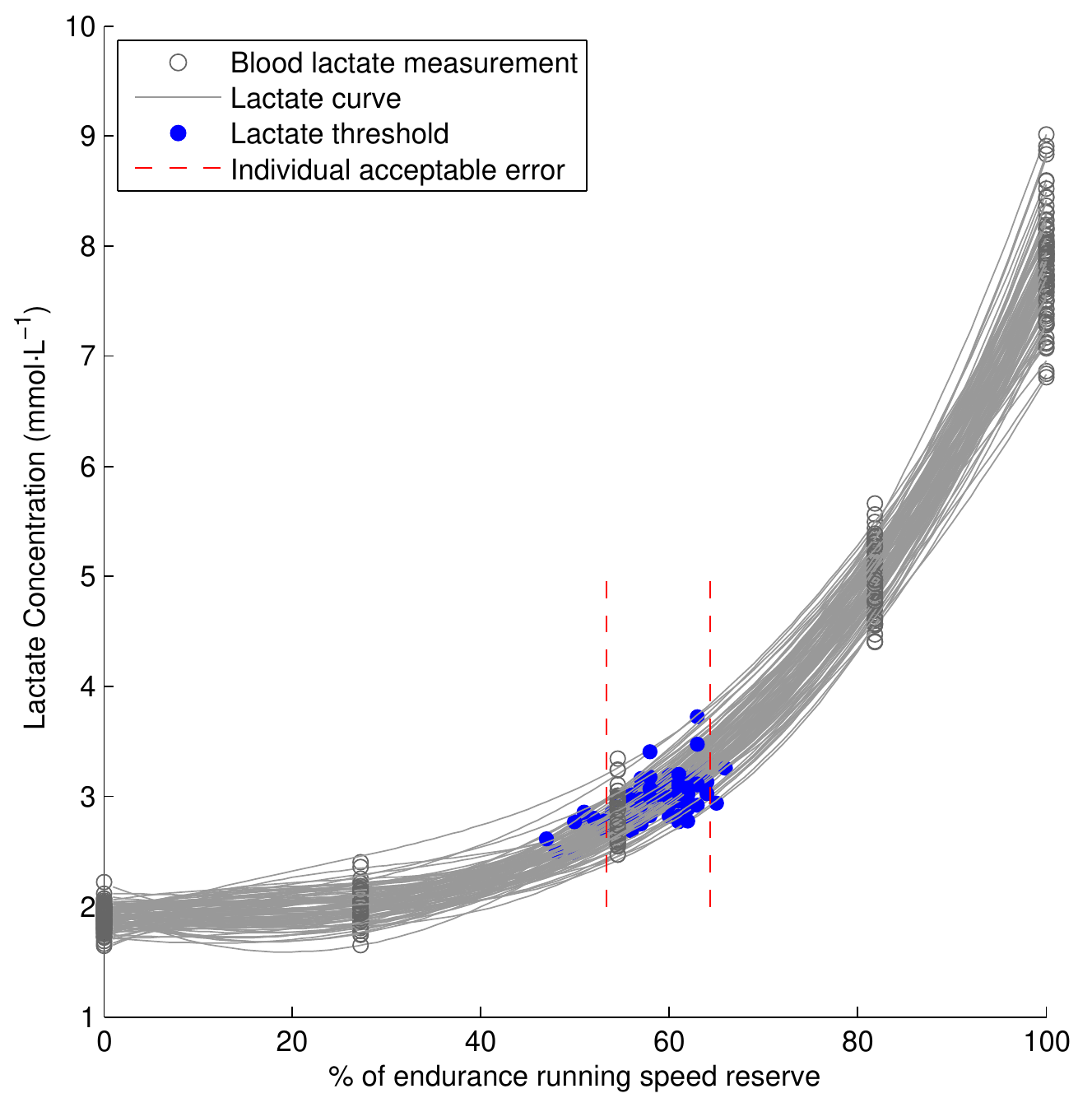}
\caption{Dmax LT variation for a given athlete caused by measurement precision error}
\label{fig:DmaxPrecision}
\end{figure}

Additionally, for each of the blood lactate measurements of each of the athletes of this population, new blood  lactate measurements (hereafter plausible blood lactate measurements) are calculated according to the precision of the measurement device \cite{Tanner2010}. In other words, these plausible blood lactate measurements correspond to the blood lactate measurements that could have been measured in the treadmill speed test due to the precision error of the blood lactate measurement device. In figure \ref{fig:DmaxPrecision} these plausible measurements for a particular athlete are represented under 'blood lactate measurements' name. Using these plausible blood lactate concentrations, their corresponding lactate curves are calculated, resulting in multiple different combinations of lactate curves that could have been derived for a particular athlete. This lactate curves are represented in figure \ref{fig:DmaxPrecision} under the 'lactate curve' name. Finally, the Dmax LT of each curve is calculated and represented in figure \ref{fig:DmaxPrecision} under the 'LT' name, which brings to light the inherent variability of the Dmax LT estimation protocol. Algorithm \ref{alg:one} is the formalization of this process.

\begin{algorithm}
\caption{Calculate error caused by measurement: Dmax protocol precision}
\label{alg:one}
\begin{algorithmic}
\REQUIRE Same number of lactate points per athlete
\STATE SDMeasurement = Blood lactate measurement precision standard deviation \cite{Tanner2010}
\STATE Precision distribution = Normal distribution(SDMeasurement)
\FOR{number of bootstrap resamples}
 \FOR{number of athletes}
  \FOR{number of lactate points} 
   \FOR{number of random samples}
	  \STATE Plausible measurement error = Random sample from Precision distribution
	  \STATE Plausible blood lactate concentration = Measured lactate + Plausible measurement error	 
   \ENDFOR
  \ENDFOR
  \STATE Plausible LTs = fDmax(Plausible blood lactate concentrations)
  \STATE Dmax error per athlete per resample = Plausible LTs - mean of Plausible LTs  	
 \ENDFOR
\ENDFOR
\STATE Dmax error distribution aggregating Dmax error per athlete per resample
\STATE Dmax protocol precision = Standard Error Measurement of Dmax error distribution

\end{algorithmic}
\end{algorithm}

In our particular case, different numbers of bootstrap resamples have been used on the sample population (10, 20 and 100). Different number of random samples have also been used for each of the blood lactate measurements (10, 20 and 100). In both cases the results from 100 random do not significantly differ from the ones obtained with 20, so a higher number of random samples is not considered necessary.

From the computation of algorithm \ref{alg:one}, we have estimated the standard deviations of the precision error of the Dmax protocol and illustrated in table \ref{tab:DmaxPrecision}, classified by the number of blood lactate points taken during the treadmill speed test (Lactate Points column, table \ref{tab:DmaxPrecision}). These results show how the precision of the Dmax LT protocol improves with the number of blood lactate measurements taken to the athletes, which is consistent with what is known in the literature and practice \cite{Santos-Concejero2014a}.

Moreover, we have also compared the variance of the Dmax LT protocol precision with the variance of the estimand used (\% of \textit{endurance running speed reserve}) in this 2$^{nd}$ iteration, so that we can evaluate how much of the variance of our estimand (the Dmax LT) can be explained by the error caused by the Dmax LT protocol. As it can be seen in table \ref{tab:DmaxPrecision}, a big amount of the variability of the Dmax LT (from 0.35 to 0.82) can be explained by the unavoidable error caused by measurement precision error.

\begin{table}
\centering
 \begin{threeparttable}
 \caption{Dmax precision according to number of lactate points}
  \begin{tabular}{@{}ccc@{}} 
  \toprule
  Lactate points & Unavoidable variability /  & variability explained  \\
								 & Dmax LT protocol precision (SD)     & by Dmax protocol precision ($R^2$) \\
  \midrule
  5 & 8.1 & 0.66\\
	6 & 5.6 & 0.74\\
  7 & 6.2 & 0.82\\
	8 & 5.3 & 0.77\\
	9 & 3.4 & 0.70\\
	10 & 2.5 & 0.35\\
  \bottomrule
  \end{tabular}
	\begin{tablenotes}
  \small
  \item Abbreviations: LT, Lactate threshold ; SD, standard deviation; It, iteration
 \end{tablenotes}	
  \label{tab:DmaxPrecision}
 \end{threeparttable}
\end{table}

\begin{figure}
\includegraphics{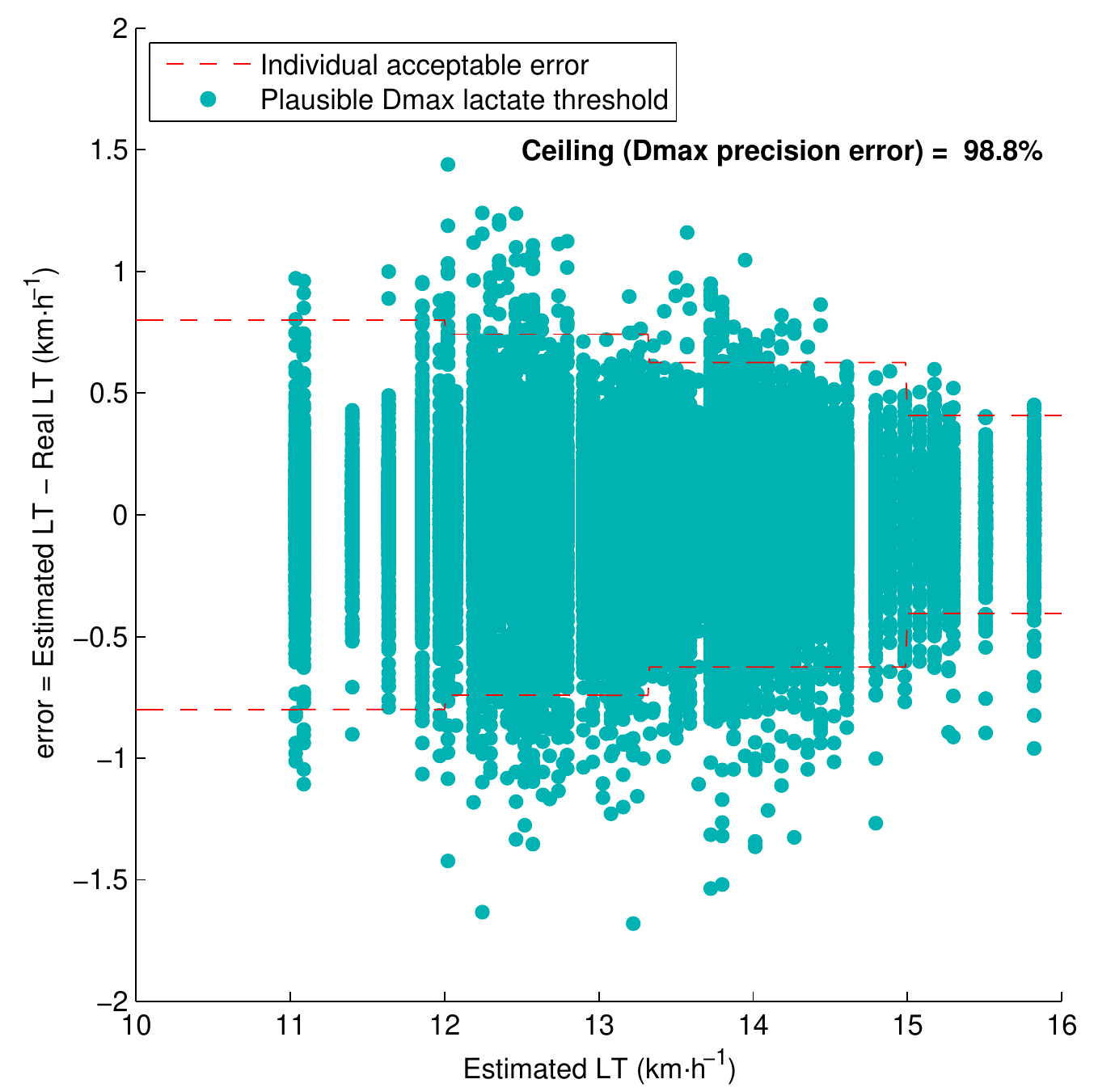}
\caption{Unavoidable error due to measurement VS Individual acceptable error}
\label{fig:DmaxPrecisionResiduals}
\end{figure}

To further analyze how the unavoidable error can affect to our estimator, we have compared it to the \textit{individual acceptable error} and represent the residual errors in figure \ref{fig:DmaxPrecisionResiduals}. In this figure, we observe that the precision error of the Dmax protocol can significantly affect the final error, showing that the 98.8\% of the plausible Dmax LTs are within the range of the \textit{individual acceptable error}. In other words, the ideal estimator created from empirical data could at best achieve a \textit{ceiling accuracy} of 98.8\%.

This methodology is an important contribution and could be easily extended to other commonly used LT estimation protocols and make a quantitative estimation of its precision. This would allow to make quantitative comparisons between protocols, something that, to the best of our knowledge, is not well addressed in the literature.

\begin{figure}
\includegraphics{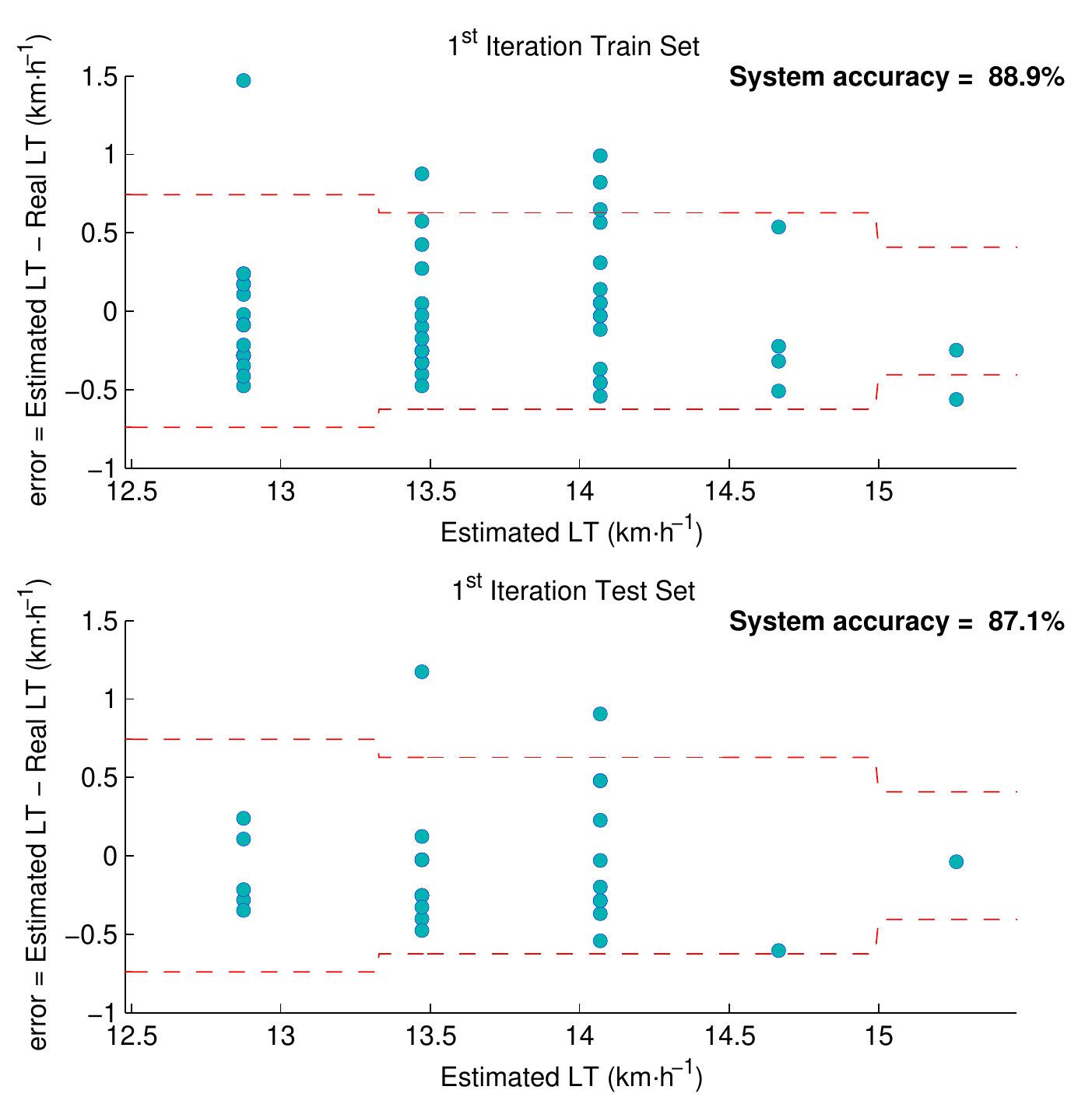}
\caption{Calibrated estimator residuals: 1$^{st}$ Iteration Data}
\label{fig:ResidualsP1}
\end{figure}

In figure \ref{fig:ResidualsP1} we represent the residuals of the \textit{calibrated estimator} with respect to the \textit{individual acceptable error} and with the 1$^{st}$ iteration data.

\begin{figure}
\includegraphics{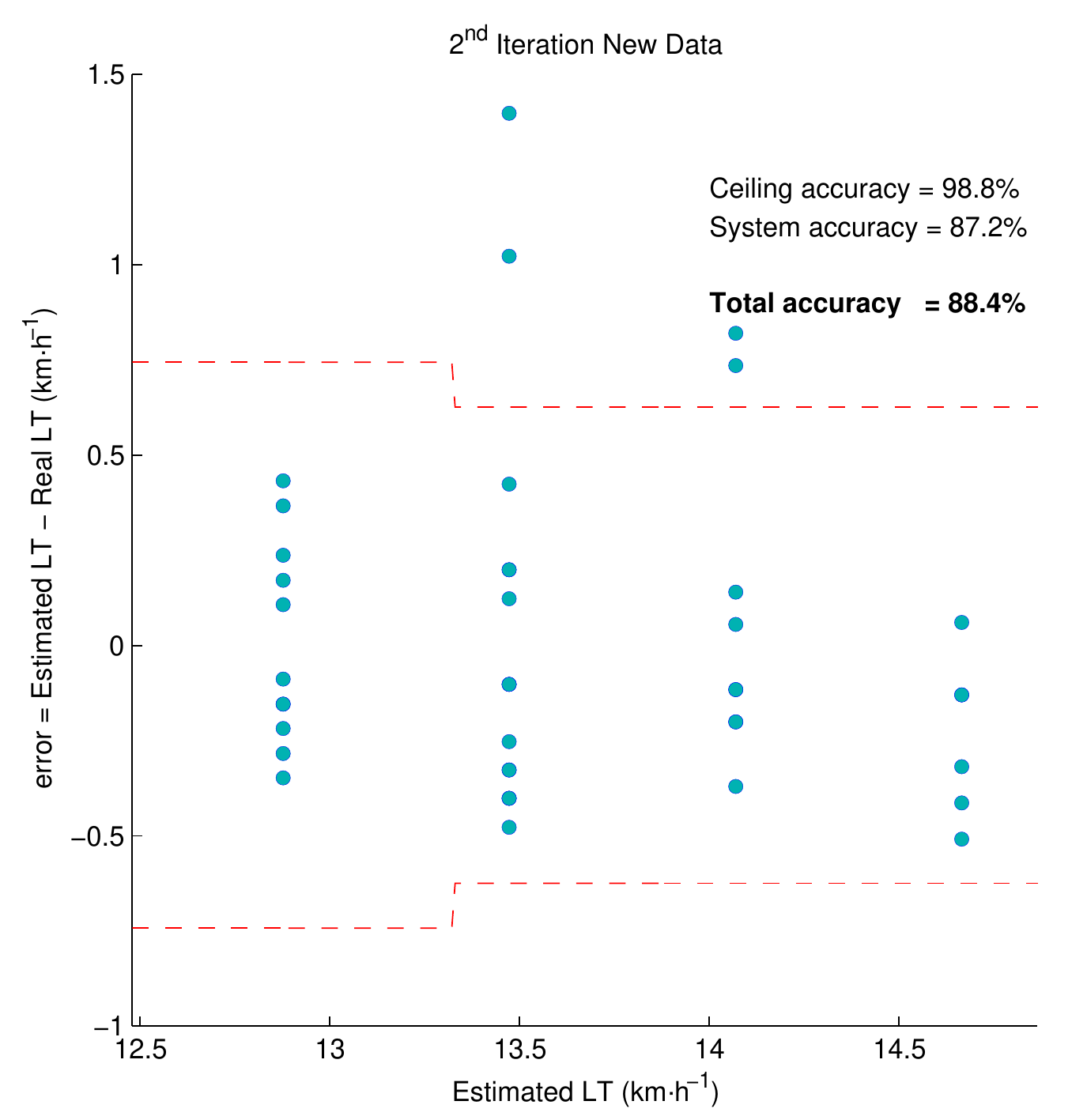}
\caption{Calibrated estimator residuals: 2$^{nd}$ Iteration New Data}
\label{fig:ResidualsP2}
\end{figure}

In figure \ref{fig:ResidualsP2} we represent the performance of the \textit{calibrated estimator} with respect to the new database and the \textit{ceiling accuracy}. The \textit{total accuracy} represents the difference between them and is used to evaluate if the system achieves the \textit{system acceptable error rate}. 

From this applicability analysis we can derive that the \textit{calibrated estimator} that we have proposed is very robust due to its high \textit{parsimony} and that the \textit{total accuracy} is very close to the \textit{system acceptable error rate}. Therefore, the calibrated estimator is proposed as a reliable and operational LT estimator.

\section{Discussion: Adjust knowledge \& next steps}
\label{sec:adjust}

In this work we have defined a set of principles necessary for estimation of the LT from empirical data and proposed an iterative strategy to implement a solution for training-decision making.

This iterative approach is inspired by strategies used in domains with high uncertainty such as software development, large engineering projects, product management, and even in sport training \cite{Jovanovic2018}. These incremental changes enable experience based learning \cite{Bar-yam2003}, due to the high uncertainty inherent of these problems, they are also suitable for machine learning. 

In this work, we have formalized and proposed an iterative approach that covers the entire process of data acquisition and pre-processing, estimator design, applicability analysis, knowledge adjustment and next steps.    

Moreover, this work reinforces the notion that appropriate problem representation (derivation of principles, feature engineering etc.) and performance evaluation (\textit{individual acceptable error}, \textit{system acceptable error rate}, comparison with the state-of-the-art etc.) combine to simplify the inference method needed to extract meaningful information from data

Additionally, this work makes several contributions to sport science. To compare the state-of-the art technique with our proposed solution, we made an analysis of the Dmax protocol which is one of the most recommended method for the LT estimation \cite{Santos-Concejero2014a}. This analysis is the first quantitative analysis of the precision error of the Dmax LT protocol in the literature, and was performed for different number of blood lactate measurements (5,6,7,8,9 and 10). More importantly, we have provided a computational method (algorithm \ref{alg:one}) that would be easily applicable to calculate the precision error of Dmax protocol with other parameters (regression function, number of points, initial speed etc.). It must be also noted that this computational method, with the appropriate adjustments, may be also useful to estimate the precision error of other LT estimation protocols.

Likewise, as represented in figure \ref{fig:DmaxPrecisionResiduals}, the unavoidable error caused by the measurement precision is sometimes even higher than the \textit{individual acceptable error}. This further validates the approximation of \textit{individual acceptable error} done in table \ref{tab:SatisfyingError} used in this work, since this constraint is even harder than those derived from the physiology equipment and methodology.

From the application perspective, we have shown that a simple heuristic (60\% of \textit{endurance speed reserve}) is capable of providing a estimation as good as the commonly used Dmax LT protocol for the recreational runners population. Unlike the Dmax LT protocol, our heuristic is an accessible solution that facilitates its consistent use as it relies on PTS: an easily measurable, non-invasive and robust feature that is well established for performance evaluation \cite{Noakes1990,Etxegarai2018,Etxegarai2018a}. Its calculation is detailed below:
\begin{enumerate}
	\item Make the treadmill speed test (detailed in section \ref{sec:Methodology}) on any standard treadmill and get the Peak Treadmill Speed (PTS)
	\item \textit{endurance running speed reserve} = PTS - Initial running speed (9 km/h in this test)
	\item LT = 60\% of \textit{endurance running speed reserve} + Initial running speed (9 km/h in this test)
\end{enumerate}

We can further compare the usefulness of this heuristic with the Dmax protocol according to the three most common use cases:  
\begin{itemize}
	\item Training prescription: The LT is widely used by coaches and athletes to aid the prescription of training intensities. 
Is in this use case where the operational nature of the heuristic here proposed stands out with respect to the Dmax LT protocol.	
	\item Prediction of performance: The LT here presented is derived from the PTS and therefore has a perfect positive correlation (R=1) with it. Hence, they both share the prediction power of PTS \cite{Noakes1990,Etxegarai2018,Etxegarai2018a} and can be interchangeably used for a qualitative prediction of performance.
	\item Training monitoring: To evaluate the physiological changes that the training has made, usually lactate curves are used in combination with the LT. Since the lactate curve also adds relevant information in this process and our solution does not provide it, the Dmax LT method its still more informative for training monitoring purposes. Despite that the use of the lactate curves are is widely extended in sport science, in this work we have shown that the lactate curves vary a lot when the blood lactate measurement precision is considered (see figure \ref{fig:DmaxPrecision}). This results suggest that it may be appropriate to make a lactate curve precision analysis to  evaluate how informative these curves really are. 
\end{itemize}

One of the possible limitations of this heuristic is that our population is drawn from local running clubs. This means that it is possible that the recreational runner population here characterized may not be representative of recreational runners of other culture, ethnicity or different contexts. However, one of the main advantages of providing a simple solution is that, unlike other black-box models, it is easily reproducible and adjustable, meaning that we have set a common ground for other researchers to evaluate the impact of our proposal. Moreover, adjusting laboratory measurements using expert knowledge and contextual information is a common practice among coaches and athletes for a successful application of the measurement. In this regard, the Dmax LT is commonly used as an initial guess. The heuristic here proposed serves the same purpose as it allows to easily adjust it (e.g. start at 60\% and decrease if the athlete is very fatigued before starting the training). In the best case scenario, future experiments done in other contexts will validate that we have been capable of discovering a common characteristic of recreational runner population. In the worst case scenario, we have provided an easy to follow methodology and an strong prior that will allow to adjust the estimator according to individual characteristics of different populations.

The main limitation of the solution presented in figure \ref{fig:ResidualsP2} may be in its generalization beyond the local recreational runners for the diverse characteristics of the worldwide recreational runners. To address this, our future work will create an adaptive framework which could be fed with other databases and allow transferability and systematization to integrate knowledge of different databases representative of worldwide populations.

\section{Conclusions}
\label{sec:conclusions}

From this work we conclude that:

From a methodological perspective, we formalize a set of clear principles for the problem in hand as well as an iterative approach that covers the entire process of data acquisition and pre-processing, estimator design, applicability analysis, knowledge adjustment and next steps. This strategy arises as a robust and adaptive approach to solve data analysis problems.

From the sport science perspective, we have set the boundaries of the Dmax LT procotol precision error and shown that other LT protocols could also be evaluated from this perspective in order to quantitatively address their reliability. Actually, we provide a computational algorithm that can serve for this purpose. 

From the application perspective, the main objective of this work was to create an accessible method to estimate the LT and to facilitate its integration into the training process of recreational runners. We have shown that a heuristic (\%60 of \textit{endurance running speed reserve}) fulfills this objective as it is as reliable as the Dmax LT protocol and covers the operational needs for a tool useful in training decision-making.

\section*{Conflicts of interest}
The authors report no conflicts of interest.

\section*{Acknowledgements}
We thank to the members of the Department of Physiology and the Department of Physical Education and Sport for their support in the problem statement and during the tests. We also thank all the athletes for participation and Mikel Echezarreta, Jon Larruskain and Ainhoa Insunza for their help in the recruitment of the athletes and providing support during the tests.
Funding: This work was supported by Grupo Campus [projects LACTATUS 2016 and LACTATUS]; Department of Economic Development and Competitiveness of the Basque Government [Gaitek 2015]; University of the Basque Country UPV/EHU [projects PPG17/56 and PPG/17/40]; Department of Education of the Basque Government [grant numbers IT914-16, US16/15 and PRE\_2015\_1\_0129]; and European Commission's Erasmus Mundus Action 2 PANTHER (Pacific Atlantic Network for Technical Higher Education and Research)” [Phd agreement No. PN/TG1/AUT/PhD/06/2017].

\section*{References}

\bibliography{EJOR18}

\end{document}